\documentstyle[12pt,epsf]{article}
\let\section=\subsection     \let\subsection=\subsubsection                

\begin{document}
\def\bbox#1{{\boldmath #1}}

\begin{center}
{\large \bf CROSS SECTIONS FOR THE DISSOCIATION OF $J/\psi$ AND
$\psi'$ BY $\pi$ AND  $\rho$ AT LOW ENERGIES }\\[2mm] 

C. Y. WONG$^1$, E. S. SWANSON$^{2,3}$ and
T. BARNES$^{1,4,5,6}$\\[5mm] 

{\small$^1$\it Physics Division, Oak Ridge National Laboratory, Oak Ridge, TN
37831 USA}\\ {\small$^2$\it Dept. of Physics and Astronomy, Univ. of
Pittsburgh, Pittsburgh, PA 15260 USA }\\ {\small$^3$\it Jefferson Lab,
Newport News, VA 23606 USA}\\

{\small
$^4$\it Department of Physics, University of Tennessee, Knoxville, TN
37996 USA}

{\small
$^5$\it Institut f\"ur Theoretische Kernphysik der Univ. 
Bonn,  Bonn, D-53115, Germany} 

{\small
$^6$\it Institut f\"ur Kernphysik, Forschungszentrum J\"ulich, 
J\"ulich, D-52425, Germany}

\end{center}

\begin{abstract}\noindent
{ Using the quark-interchange model of Barnes and Swanson, we evaluate
the dissociation cross sections for $J/\psi$ and $\psi'$ by $\pi$ and
$\rho$ at low collision energies. In collisions with $\pi$ mesons near
threshold the $\pi + J/\psi$ dissociation cross section is predicted
to be small, but the $\pi +\psi'$ cross section is found to be quite
large, in qualitative agreement with experimental data on $J/\psi$ and
$\psi'$ production in high-energy heavy-ion collisions.  The $\rho +
J/\psi$ and $\rho + \psi'$ dissociation cross sections are both
predicted to be large numerically and are divergent near threshold.
These $\rho + J/\psi$ and $\rho + \psi'$ cross sections show
considerable sensitivity to the mass of the $\rho$ meson if it is off
energy shell.  }
\end{abstract}

\section{Introduction}

In a high-energy heavy-ion collision, the fate of a $J/\psi$ or
$\psi'$ produced in the collision depends sensitively on its
dissociation cross sections with ordinary light hadrons.  If we are to
use $J/\psi$ production as an indicator of the formation of a
quark-gluon plasma in a high-energy heavy-ion collision, as suggested
by Matsui and Satz \cite{Mat86}, we should concurrently evaluate the
competing process of $J/\psi$ absorption by hadrons that are produced
during the collision 
[2-10].

The dissociation of a $J/\psi$ by light mesons has been examined
previously, but the cross sections from various analyses unfortunately
differ considerably, due largely to different assumptions regarding
the dominant scattering mechanism 
[11-18].
A description of the present calculations is found in
Ref. \cite{Won99a}.

Charmonium dissociation processes can presumably be described in terms
of the fundamental quark and gluon interactions but are of greatest
phenomenological interest at energy scales in the resonance region.
For this reason, we advocate the use of the known quark-gluon forces to
specify the underlying scattering amplitude, which must then be
convolved with the explicit nonrelativistic quark model hadron
wavefunctions of the initial and final mesons to obtain the
dissociation cross section.

\section{The model}

The model of Barnes and Swanson, which we shall use to describe these
processes, is a quark interchange model, with a quark-quark interaction
taken from quark potential models \cite{Bar92}.  In this model the
meson-meson scattering amplitude is given by the sum of four quark
line diagrams, which are shown in Fig.~1. These are evaluated as
overlap integrals of quark model wavefunctions, using the ``Feynman
rules" given in App.~C of Ref. \cite{Bar92}.  This method has
previously been applied successfully to the closely related
no-annihilation scattering channels $I=2$ $\pi\pi$ \cite{Bar92},
$I=3/2$ $K\pi$ \cite {Kpi}, $I=0,1$ S-wave $KN$ scattering \cite{KN},
and the short-range repulsive NN interaction \cite{NN}.

\vspace*{4.5cm}
\epsfxsize=300pt
\includegraphics{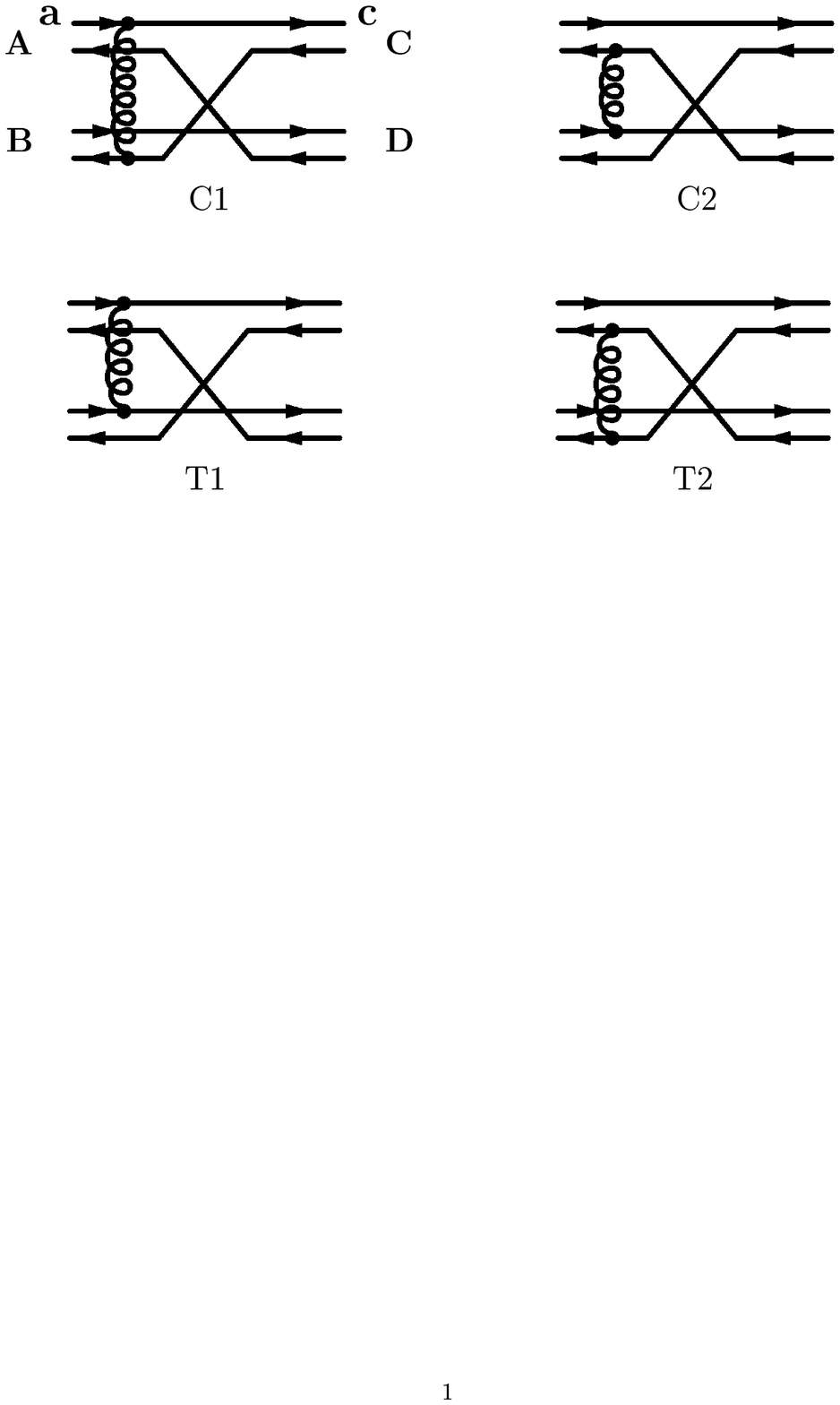}
\vspace*{0.4cm}\hspace*{1.5cm}
\begin{minipage}[t]{14cm}
\noindent {\bf Fig.\ 1}.  {Quark-exchange diagrams included in the
calculation.}
\end{minipage}
\vskip 4truemm
\noindent 

The diagrams of Fig.\ 1 are the `prior' forms; there are also `post'
forms, in which the interaction takes place after the interchange of
the quarks.  In the non-relativistic formulation one can show that
these two forms are equivalent \cite{Sch68,Swa92}, although we find
small differences between them when we use relativistic kinematics for
the initial and final mesons.

We take the interaction $H_{ij}(r_{ij})$ between
quarks $i$ and $j$, represented by the curly line in Fig.\
1, to be the conventional Coulomb plus linear plus spin-spin hyperfine 
interaction,
\begin{eqnarray}
\label{eq:Hij}
H_{ij}(r_{ij})
={\bbox{\lambda}(i) \over 2}\cdot {\bbox{\lambda}(j) \over 2} \left \{
{\alpha_s \over r_{ij}} - {3 b \over 4} r_{ij} - {8 \pi \alpha_s \over
3 m_i m_j } \bbox{S}_i \cdot \bbox{S} _j \; {\sigma^3 \over
\pi^{3/2} }\,  e^{-\sigma^2 r_{ij}^2}  
+ V_{con}
\right
\},
\end{eqnarray}
where $\alpha_s$ is the strong coupling constant, $b$ is the string
tension, $m_i$ and $m_j$ are the masses of the interacting
constituents, and $\sigma$ is the range parameter in the
Gaussian-regularized contact hyperfine term.  (For antiquarks, the
generator $\bbox{\lambda}/2$ is, as usual, replaced by
$-\bbox{\lambda}^{T}/2$.)  We used the parameter set
\vspace*{-0.1cm}
\begin{eqnarray}
\label{eq:par}
\alpha_s&=&0.58,~~~b=0.18\ {\rm GeV}^2,~~~\sigma=0.897 {\rm~GeV}, 
\nonumber\\
m_u&=&m_d=0.345 {\rm ~ GeV}, ~~~m_c=1.931 {\rm ~GeV},
~~V_{con}=-0.612 {\rm ~GeV},
\end{eqnarray} 
which gives a reasonably accurate description of the $q\bar q$ meson
spectrum and, when applied to scattering, also gives an $I=2$ $\pi\pi$
phase shift, which is in good agreement with experiment.  This
Hamiltonian was used in the nonrelativistic Schr\"odinger equation to
generate $q\bar q$ wavefunctions, which were in turn used with the
same Hamiltonian in the diagrams of Fig. 1 to evaluate the scattering
amplitudes and cross sections.  We also used a second set of
parameters, $\alpha_s = 0.594$, $b = 0.162$ GeV$^2$, $\sigma = 0.897$
GeV, $m_u = m_d = 0.335$ GeV, $m_c = 1.6$~GeV and a
flavor-dependent $V_{con}$, found by fitting a large set of
experimental masses, to test the sensitivity of our results to
parameter variations.

\vspace*{-0.4cm}
\section{Cross section for the dissociation of $J/\psi$ and $\psi'$ by $\pi$}

\vspace*{-0.1cm}
The dissociation of $J/\psi$ and $\psi'$ by pions of sufficient energy
can lead to many different open-charm channels.  In $\pi + J/\psi$ and
$\pi + \psi'$ collisions the channels with the lowest thresholds are
$D \bar {D^*} $, $D^* {\bar D}$, and $ D^* \bar {D^*}$.

\vspace*{-0.9cm}
\epsfxsize=300pt
\includegraphics{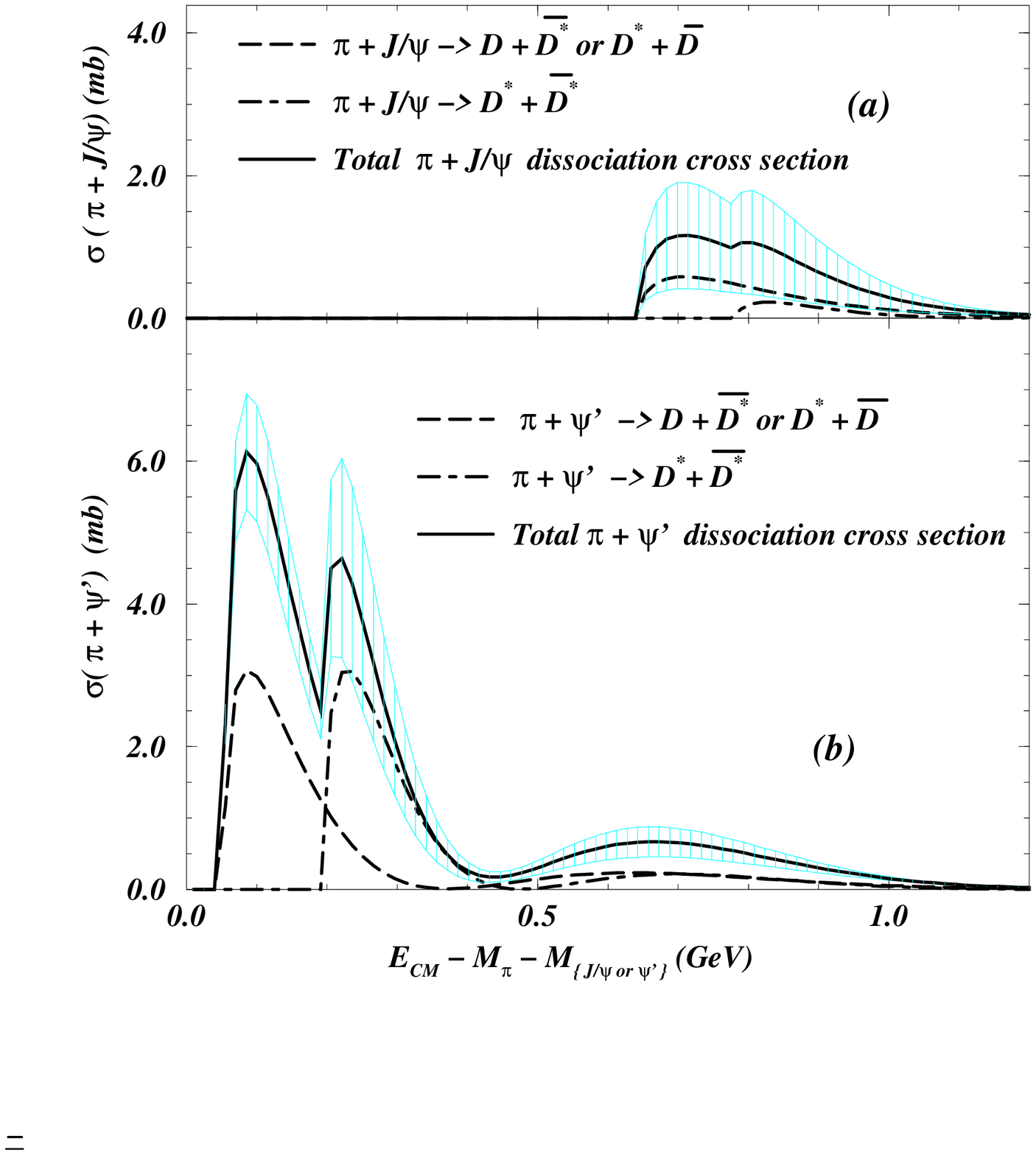}
\vspace*{10.1cm}\hspace*{1.0cm}
\begin{minipage}[t]{11cm}
\noindent {Fig.\ 2.  Total and partial $\pi + J/\psi$ (Fig.~2$a$) and
$\pi + \psi'$ (Fig.~2$b$) dissociation cross sections.}
\end{minipage}

\newpage
In Fig.\ 2 we show the cross sections for the dissociation of $J/\psi$
and $\psi'$ by $\pi$ as a function of the initial kinetic energy in
the center-of mass system, $E_{KE}=\sqrt{s}-m_A-m_B$, where $A$ and
$B$ are the colliding particles.  The total cross sections are shown
as thick solid curves and are the means of the `prior' and `post'
results.  The estimated errors due to the post-prior discrepancy and
parameter variations are indicated by bands in the figures.  The total
cross section for the dissociation of $J/\psi$ by $\pi$ has an initial
kinetic energy threshold of 0.65 GeV, and is about 1 mb just above
threshold (Fig.\ 2$a$).  This $\pi+J/\psi$ cross section is rather
smaller than the 7 mb obtained by Martins $et~al.$ \cite{Mar95}, who
used the same formalism but made different assumptions about the
confining interaction.

The threshold for $\pi+\psi'\rightarrow\{ D\bar {D^*},D^*{\bar D}\} $
is only 0.05 GeV.  The total cross section for the dissociation of
$\psi'$ by $\pi$ has local maxima at $E_{KE}\sim 0.1$ GeV and $\sim
0.22$ GeV, where the cross section is predicted to be, respectively,
6.2(0.8) mb and 4.6(1.8) mb (Fig.\ 2$b$).  Thus, we find that the low
energy $\pi + J/\psi$ dissociation cross section should be rather
small, but in contrast we predict the $\pi + \psi'$ cross section to
be quite large. This is in qualitative agreement with earlier analyses
of $J/\psi$ and $\psi'$ suppression in pA, O-A, and S-U collisions
\cite{Won96,Won98}.

\section{Cross section for the dissociation of $J/\psi$ and $\psi'$ by $\rho$}

We next calculate the dissociation cross sections for $J/\psi$ and
$\psi'$ in their collisions with $\rho$ mesons.  A $\rho$ meson is a spin-1
particle, so the total spin of $\rho + \{ J/\psi {\rm ~or~} \psi'\}$ is
$S_{tot}=0,1,$ and $2$. 
The lowest mass final reaction products can be $( D,\bar D )$ with
$(S_3,S_4)=(0,0)$ and $S_{tot}=0$, $(D,\bar{D^*} )$
with $(S_3,S_4)=(0,1)$ or $( {D^*},\bar D)$ with $(S_3,S_4)=(1,0)$
for $S_{tot}=1$, and $({D^*},\bar D^* )$
with $(S_3,S_4)=(1,1)$, for which $S_{tot}$ can be 0, 1 or 2.

Note that since the reaction $\rho+J/\psi \to D \bar D$ is exothermic,
{\it this cross section diverges as} $1/|\vec v_{\rho - (J/\psi) }|$ {\it
near threshold, independent of the detailed assumptions regarding the
scattering mechanism}.  For the other channels the thresholds lie
above $m_\rho + m_{J/\psi}$, so those subprocesses are endothermic and
have zero cross section at threshold.  The total dissociation cross
section we find is shown as a solid line in Fig. 3.  Numerically it is
quite large, about 11(3) mb for an initial-state kinetic energy of 0.1
GeV, decreasing to 6(2) mb at a kinetic energy of 0.2 GeV.

\newpage

\vspace*{-1.5cm}
\epsfxsize=300pt
\includegraphics{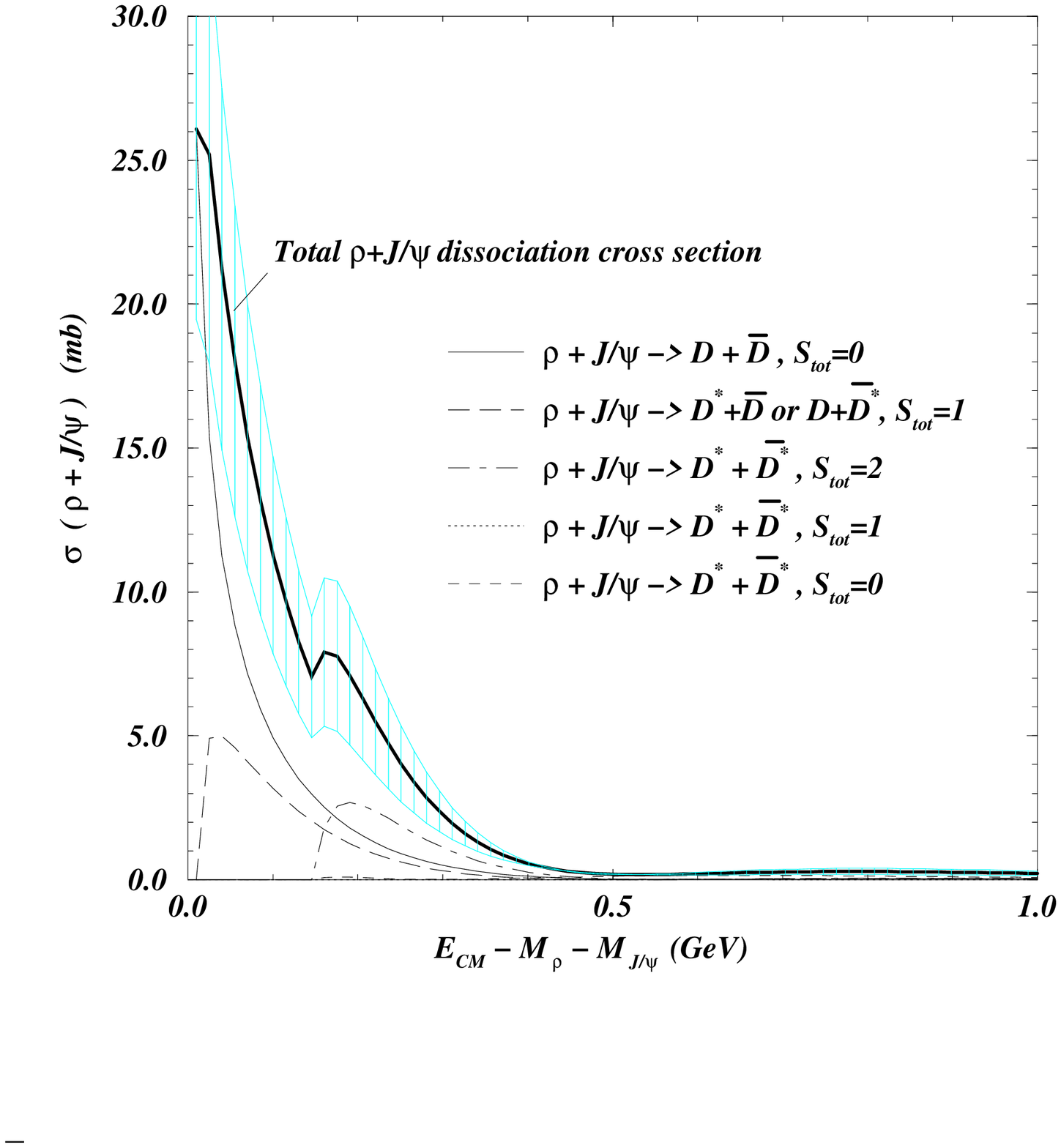}
\vspace*{9.8cm}\hspace*{1.5cm}
\begin{minipage}[t]{10cm}
\noindent {Fig.\ 3.  Cross sections for the dissociation of
$J/\psi$  by $\rho$ into the six lowest open-charm channels.
}
\end{minipage}
\vskip 4truemm
\noindent 

\section{Dependence of the dissociation cross sections on the 
$\rho$ mass}

The collision of two pions leads to isospin $I=0,$ 1 and 2 states.
The $I=1$ cross section at low energies is dominated by the
$\rho(770)$, which has a width of 0.150 GeV.  Because of this large
width, $\rho$ mesons can be formed significantly far off the energy
shell in $\pi\pi$ collisions.

A $J/\psi$ interacting with off-shell $\rho$ mesons will have a range
of different effective thresholds, so it is useful to calculate these
dissociation cross sections as a function of the $\rho$ meson mass.
For this calculation we assume that the off-shell $\rho$ has the same
spatial wavefunction as on-shell, and that only the kinematics of the
reaction change.  The results are shown in Fig. 4 for the $J/\psi$
dissociation cross section as a function of the center-of-mass kinetic
energy above the lowest threshold, $E_{KE}'$. Here,
$E_{KE}'=\sqrt{s}-m_\rho-m_{J/\psi}-E_{th}(0)$, where $E_{th}(0)$ is
the lowest threshold, and is given by $(m_D+m_{\bar
D}-m_{J/\psi}-m_\rho )\Theta(m_D+m_{\bar D}-m_{J/\psi}-m_\rho)$.

Figure 4 shows that the cross section increases from about 1 mb to 3
mb as $m_\rho$ increases from 0.45 GeV to 0.55 GeV.  At $m_\rho=0.65$
GeV, the reaction $\rho$+$J/\psi$ becomes exothermic, and the cross
section completely changes character; it diverges as
$1/\sqrt{E_{KE}'}$ near $E_{KE}'=0$.

The results in Fig.\ 4 show that the dissociation cross section for
$\rho + J/\psi$ is quite sensitive to the $\rho$ mass; higher mass
$\rho$ mesons have much larger cross sections near threshold.

\vspace*{-0.5cm}
\epsfxsize=300pt
\includegraphics{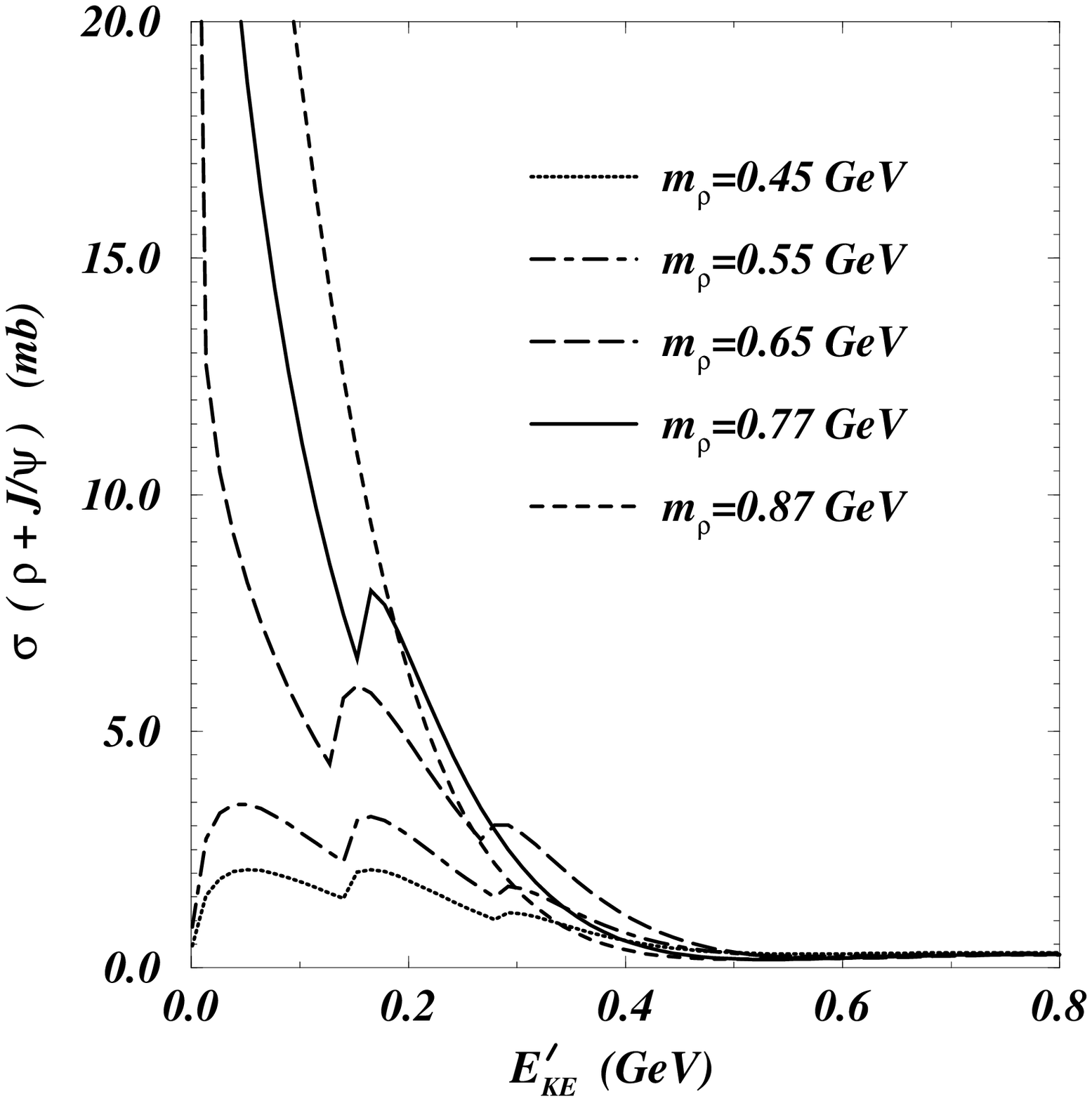}
\vspace*{9.8cm}\hspace*{1.5cm}
\begin{minipage}[t]{11cm}
\noindent {Fig.\ 4.  The total $\rho + J/\psi$ dissociation cross section
for different values of
$m_\rho$.
}
\end{minipage}
\vskip 4truemm
\noindent 

We have carried out the corresponding calculations for the $\rho +
\psi'$ dissociation cross sections as a function of the $\rho$ meson
mass; the results are shown in Fig. 5 as a function of
$E_{KE}'=\sqrt{s}-m_\rho-m_{\psi'}-E_{th}(0)$, where the lowest
threshold is $E_{th}(0)=(m_D+m_{\bar
D}-m_{\psi'}-m_\rho)\Theta(m_D+m_{\bar D}-m_{\psi'}-m_\rho)$.  The
cross section diverges as $1/\sqrt{E_{KE}'}$ near $E_{KE}' = 0$ for
the $\rho$ masses considered in Fig.\ 5.  There are additional
contributions from the $D^* \bar D$ and $D^* \bar D^*$ channels.  For
a given $E_{KE}'$ and $m_\rho$, the $\psi'$ dissociation cross section
is larger than the $J/\psi$ cross section, with the exception of
$m_\rho=770$ MeV and $E_{KE}'>0.2$ GeV, at which the two cross
sections are nearly equal.

The results of Fig.\ 5 show that the dissociation cross section
for $\psi'$ by an off energy shell $\rho$ is quite sensitive to 
$m_\rho$.  

\vspace*{-0.5cm}
\epsfxsize=300pt
\includegraphics{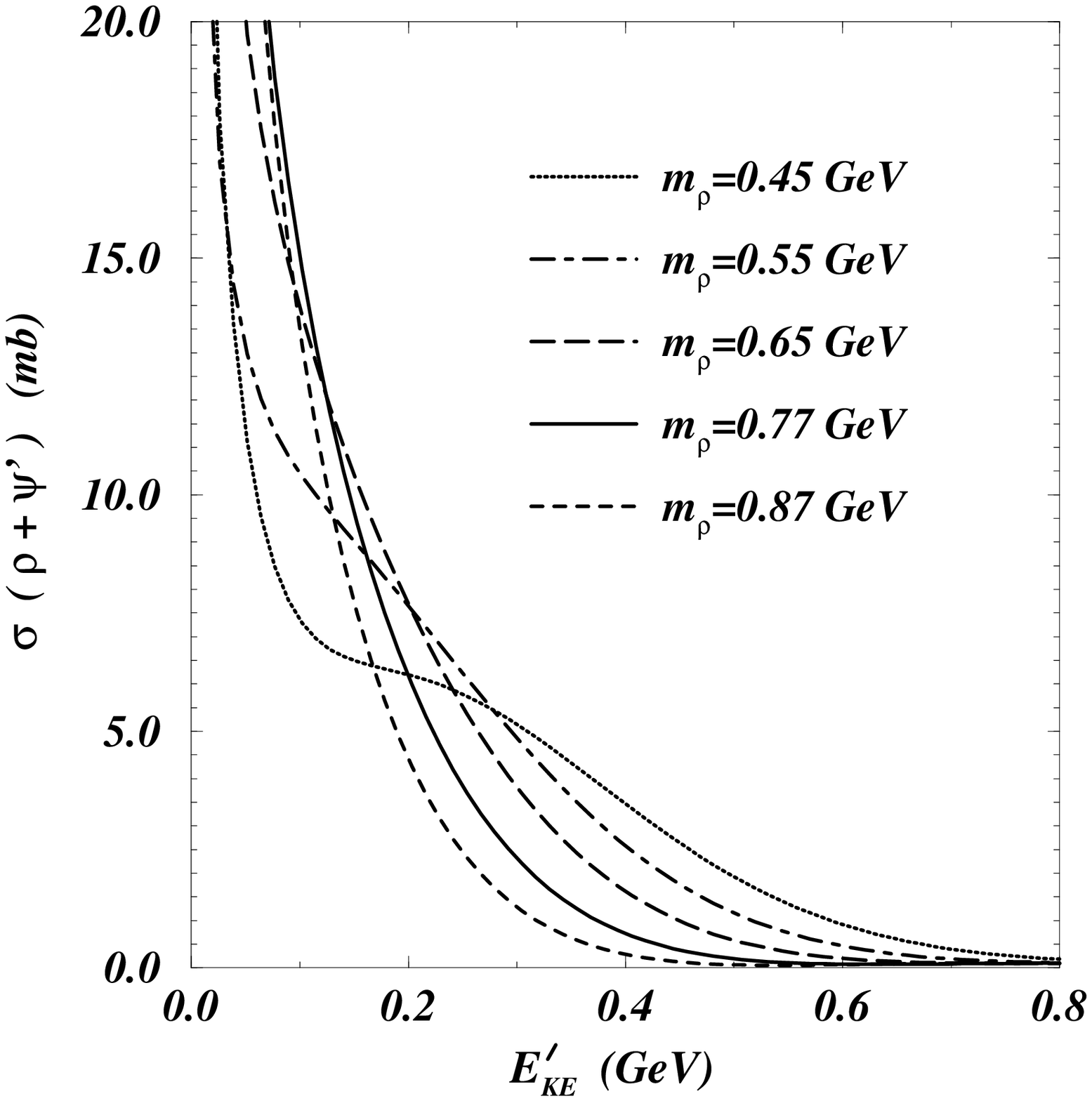}
\vspace*{9.8cm}\hspace*{1.5cm}
\begin{minipage}[t]{11cm}
\noindent {Fig.\ 5.  The total dissociation cross sections for 
$\psi'$  by $\rho$ for different $m_\rho$.
}
\end{minipage}
\vskip 4truemm
\noindent 
\vspace*{-0.7cm}

\section{Conclusions and future applications}

The quark-interchange scattering model of Barnes and Swanson has been
supported by extensive comparisons with experimental data in channels
such as $I=2$ $\pi\pi$ scattering, $I=3/2$ $K\pi$ scattering, $I=0,1$
S-wave $KN$ scattering, the small $\pi + J/\psi$ and large
$\pi$+$\psi'$ dissociation cross sections, and indirectly through its
incorporation of a quark model Hamiltonian that gives a good
description of the experimental meson spectrum.  The results for $\rho
+ J/\psi$ and $\rho + \psi'$ dissociation cross sections we have found
using this model are presumably qualitatively correct, since they
depend mainly on the endothermic or exothermic nature of these
processes. The more detailed numerical predictions certainly need to
confront experimental data through a detailed study of $J/\psi$ and
$\psi'$ production and evolution in high-energy heavy-ion collisions.

It may also be possible to infer the $\rho$ dissociation cross
sections from studies of heavy-ion collisions.  $\rho$ mesons are
produced in $\pi\pi$ collisions and they also decay inversely into two
pions.  The production and the decay of $\rho$ mesons can readily
reach an equilibrium and the the law of mass action applies.  As a
consequence, the density of $\rho$ mesons should vary as the square of
the pion density.  In a heavy-ion collision the density of pions
produced varies approximately as the path length $L$ of the colliding
nuclei \cite{Won96}.  Thus, the density of $\rho$ mesons depends
approximately quadratically on the path length $L$.  Charmonium
absorption by these $\rho$ mesons thus leads to a term $c
\times\sigma(\rho + J/\psi)\times L^2$ in the exponential absorption
index.  This quadratic dependence will be more evident if the cross
section $\sigma(\rho$+$J/\psi)$ for $\rho + J/\psi$ dissociation is
large, as we have found here.  This term will also be enhanced as we
increase the radii of the colliding nuclei, for example in going from
S+U to Pb+Pb collisions.  The degree to which the observed anomalous
suppression in Pb+Pb collisions is due to $\rho + J/\psi$ dissociation
will require a careful quantitative study, taking into account the
variation of the dissociation cross section on the off energy shell
$\rho$ mass.  In any case, the possible importance of such a nonlinear
suppression term should be considered in the attempt to identify the
creation of a quark-gluon plasma through suppression of $J/\psi$
production.

\vspace*{-0.3cm}
\section*{Acknowledgments}

This research was supported by the Division of Nuclear Physics, DOE,
under Contract No. DE-AC05-96OR21400 managed by Lockheed Martin Energy
Research Corp. ES acknowledges support from the DOE under grant
DE-FG02-96ER40944 and DOE contract DE-AC05-84ER40150 under which the
Southeastern Universities Research Association operates the Jefferson
National Accelerator Facility. TB acknowledges additional support from
the Deutsche Forschungsgemeinschaft DFG under contract Bo 56/153-1.
The authors would also like to thank C. M. Ko and S. Sorensen for
useful discussions.


\vspace*{-0.3cm}
\begin{thebibliography}{99}

\bibitem{Mat86} T. Matsui and H. Satz, Phys. Lett. {\bf B178} (1986) 416.

\bibitem{Gon96} M. Gonin, NA50 Collaboration, Nucl. Phys. 
A610 (1996) 404c.

\bibitem{Rom98} A. Romana $et~al.$, NA50 Collaboration, in Proceedings
of the XXXIII Recontres de Moriond, Les Arcs, France, 21-28 March,
1998.

\bibitem{Won96} C. Y. Wong, Nucl. Phys.  {\bf A610} (1996) 434c;
C. Y. Wong, Nucl. Phys. {\bf A630} (1998) 487. 

\bibitem{Won98} C. Y. Wong, hep-ph/9809497.  Proceedings of Workshop
on Charmonium Production in Relativistic Nuclear Collisions, Institute
of Nuclear Theory, Seattle, May 18-22, 1998.

\bibitem{Kha96} D. Kharzeev, Nucl. Phys.  {\bf A610} (1996) 418c;
D. Kharzeev, Nucl. Phys. A638  (1998) 279c.

\bibitem{Bla96} J.-P. Blazoit and J.-Y. Ollitrault, 
Nucl. Phys.  {\bf A610} (1996) 452c.

\bibitem{Cap96} A. Capella, A. Kaidalov, A. K. Akil, and C. Gerschel,
Phys. Lett. {\bf B393} (1997) 431.

\bibitem{Cas96} W. Cassing and C. M. Ko, Phys. Lett. {\bf B396} (1997) 39;
 W. Cassing, E. L. Bratkovskaya, Nucl. Phys. {\bf A623} (1997) 570.

\bibitem{Sa99}
Sa Ben-Hao $et~al.$, J. Phys. {\bf G25}  (1999) 1123.

\bibitem{Kha94} D. Kharzeev and H. Satz, Phys. Lett. {\bf B334}
(1994) 155.

\bibitem{Kha96a} D. Kharzeev, H. Satz, A. Syamtomov, and G. Zinovjev,
Phys. Lett. {\bf B389} (1996) 595.

\bibitem{Mat98}
S. G. Matinyan and B. M\"uller, Phys. Rev. {\bf C58} (1998) 2994.

\bibitem{Hag99}
K. L. Haglin, nucl-th/9907034.


\bibitem{Mar95}
K. Martins, D. Blaschke, and E.\ Quack, Phys.\ Rev. {\bf C51} (1995) 2723.

\bibitem{Won99a}
C. Y. Wong, E. S. Swanson, and T. Barnes, hep-ph/99120431.

\bibitem{Lin99}
Z. W. Lin and C. M. Ko, nucl-th/9912046.

\bibitem{Bla00}
D. B. Blaschke, G. R. G. Burau, M. A. Ivanov,
Yu. L. Kalinovsky, and P. C. Tandy, hep-ph/0002047.


\bibitem{Bar92}
T. Barnes and E. S. Swanson, Phys. Rev. {\bf D46} (1992) 131.


\bibitem{Kpi}
T. Barnes, E. S. Swanson and J. Weinstein, Phys. Rev. {\bf D46} (1992) 4868.

\bibitem{KN}
T. Barnes and E. S. Swanson, Phys. Rev. {\bf C49} (1994) 1166.

\bibitem{NN} T. Barnes, S. Capstick, M. D. Kovarik and E. S. Swanson,
Phys. Rev. {\bf C48} (1993) 539.

\bibitem{Sch68}
L. I. Schiff, {\it Quantum Mechanics} (McGraw-Hill, New York, 1968), 
pp. 384-387. 

\bibitem{Swa92}
E. S. Swanson, Ann. Phys. (N.Y.) { \bf 220} (1992) 73.

\end{thebibliography}
\end{document}